\documentclass[aps,prl,nobibnotes,nofootinbib,preprintnumbers,amsmath,amssymb,latexsym,array,enumerate,letter,twocolumn,superscriptaddress]{revtex4}
\usepackage{mathptmx}%
\usepackage{bbm}
\usepackage{bm}
\usepackage{amsfonts}
\usepackage{amsthm}
\usepackage{graphicx}
\usepackage[english]{babel}
\usepackage{multirow}
\usepackage{float}
\usepackage{url}
\usepackage{slashed}
\usepackage{xcolor} 
\usepackage[utf8]{inputenc}
\usepackage{stmaryrd} 
\usepackage{enumitem}
\usepackage{hyperref}
\usepackage{verbatim}
\usepackage{ulem}
\theoremstyle{remark}

\newcommand{\eV}{\rm{eV}}
\newcommand{\keV}{\rm{keV}}
\newcommand{\MeV}{\rm{MeV}}
\newcommand{\GeV}{\rm{GeV}}

\begin{document}

\title{\boldmath Axion-assisted Resonance Oscillation Rescues the Dodelson-Widrow Mechanism}

\author{Shu-Yuan Guo }
\email{shyuanguo@ytu.edu.cn}
\author{Xuewen Liu }
\email{xuewenliu@ytu.edu.cn}
\author{Bin Zhu}
\email{zhubin@mail.nankai.edu.cn}

% The "\note" macro will give a warning: "Ignoring empty anchor..."
% you can safely ignore it.

\affiliation{Department of Physics, Yantai University, Yantai 264005, China}

% e-mail addresses: one for each author, in the same order as the authors

\begin{abstract}
The $\keV$ scale sterile neutrino was a qualified candidate for dark matter particles in the Dodelson-Widrow mechanism. But the mixing angle, needed to provide enough amount of dark matter, is in contradiction with the astrophysical observations. To alleviate such tension, we introduce an effective interaction, i.e. $g_a (\phi/\Lambda)\partial_{\mu}a \overline{\nu_\alpha}\gamma^{\mu} \gamma_5 \nu_\alpha$, among Standard Model neutrino $\nu_\alpha$, axion $a$, and singlet $\phi$. The axial-vector interaction form is determined by the axion shift symmetry, 
and the singlet $\phi$ with dynamically varied vacuum expectation value is introduced to reinforce the axial-vector coupling strength and evade the stringent neutrino oscillation constraints. The effective potential generated by the new interaction {could cancel} the SM counterpart, resulting in an {enhanced converting} probability between SM neutrino and sterile neutrino. Hence, the production rate of sterile neutrinos can be substantially enlarged with smaller mixing compared to the DW mechanism.

\end{abstract}

\maketitle

%%%%%%%%%%%%%%%%%%%%%%%%%%%%%%%%%%
\section{Introduction}
%%%%%%%%%%%%%%%%%%%%%%%%%%%%%%%%%%
	
The mystery surrounding dark matter (DM) production and neutrino mass generation is intriguing. The seesaw mechanism,  which introduces an extra neutrino $\nu_s$, is the minimal and natural framework for solving these two issues.  After the electroweak symmetry breaking, the Dirac mass terms $yv \bar{\nu}_\alpha \nu_s + \rm{h.c.}$ will inevitably result in a small mixing between the active and sterile neutrinos, $\nu_4=\cos \theta \nu_s+\sin \theta \nu_\alpha$, 	which makes the sterile neutrino an ideal dark matter candidate in the Dodelson-Widrow (DW) mechanism~\cite{Dodelson:1993je}. This mass mixing $\theta$ is too tiny in the early universe to thermalize the sterile neutrinos with the Standard Model (SM) bath. Thus there were no initial distributions for sterile neutrino. However, by the collisions of active neutrinos with other SM particles, even a slight amount of mixing can produce a sizable population of sterile neutrinos. The production process will end once the SM neutrinos decoupled from the thermal bath and the relic abundance of sterile neutrinos is attained with $\theta\sim 10^{-6}$ and $m_{\nu_s}\sim 100\mathrm{keV}$~\cite{Abazajian:2017tcc}. Unfortunately the active-sterile neutrino mixing is already in tension with searches for DM decaying into monochromatic X-rays in this minimal setup~\cite{Watson:2011dw,Horiuchi:2013noa,Perez:2016tcq,Dessert:2018qih,Ng:2019gch}. 
Together with the bounds from the observation on the dwarf galaxies~\cite{Tremaine:1979we,Boyarsky:2008ju,Merle:2015vzu,Abazajian:2017tcc}, the parameter space has been almost ruled out.

Given these challenges, less minimal hypotheses that would increase the rate of sterile neutrinos and reduce the restrictions imposed by astrophysical observations have been proposed. The existence of a significant lepton-number asymmetry in the early universe to enlarge the sterile neutrino production rate resonantly is a common new physics approach~\cite{Shi:1998km} to alleviate the above tension. The problem with this approach is that it can not explain why the lepton asymmetry is significantly larger than the baryon asymmetry, and get a strong constraint from Big Bang Nucleosynthesis.  Including additional self-interaction between active neutrinos is another elegant approach~\cite{DeGouvea:2019wpf,Kelly:2020pcy,Kelly:2020aks} to erasing the tension. It helps efficiently produce sterile neutrino DM and keeps neutrinos in thermal equilibrium with themselves for a longer time. The neutrinophilic scalar or vector is responsible for mediating the self-interaction, which results in a connection between production rate and dark radiation $\Delta N_{\mathrm{eff}}$. A similar approach to generating the appropriate relic abundance of $\nu_s$ is the freeze-in production via Higgs singlet decay or vector boson decay processes~\cite{Kusenko:2006rh,Petraki:2007gq,Roland:2014vba,Drewes:2016upu,Shuve:2014doa}. 
Ref.~\cite{Berlin:2016bdv} proposed a solution by introducing interactions with axion-like particles, which breaks the connection of the mixing angles that responsible for sterile neutrino production in the early universe and for the decay of sterile neutrino at late times. 
The size of the mixing decreases as the energy density of axion-like particles diluted over the cosmological expansion, thus avoiding the X-ray bounds.
Even though 
all above proposals can explain the sterile neutrino relic abundance, they go far beyond the minimal configuration. The resonant oscillation seems to be the best option if we insist on preserving the grace of the DW mechanism. 

It is demonstrated that the presence of a primordial condensate of the dark photon can alter the dispersion relation of the neutrinos in the early universe~\cite{Alonso-Alvarez:2021pgy}. A greater active neutrino self-energy allows for a level crossover with its heavier sterile counterpart, enabling resonant oscillations between the two states. We notice that the underlying principle of the approach is that a dark photon originates from the misalignment mechanism where it has to couple to the curvature {\it R} with extra substantial, highly tuned couplings. But these couplings trigger perturbative unitarity violations in longitudinal photon-graviton scattering at low energies\cite{Arias:2012az}. This straightforward and elegant approach also faces problems.  

We point out that the axion can be used to generate resonant oscillation, as it naturally originates from the misalignment mechanism\cite{Linde:1987bx,Preskill:1982cy}. The minimal interaction between axion and active neutrino $g_a\partial_{\mu}a \overline{\nu_\alpha}\gamma^{\mu}\gamma_5 \nu_\alpha$ suffers from stringent constraints, where the magnitude of $g_a$ required for resonance to exist contradicts the neutrino oscillation in the current universe. In this letter, we propose a brand-new method to generate resonant oscillations in the early universe and remove the neutrino oscillation constraints in the late universe through the axion assisted DW mechanism (denoted as aDW hereafter), with interaction $g_a(\phi/\Lambda)\partial_{\mu}a \overline{\nu_\alpha}\gamma^{\mu} \gamma_5\nu_\alpha$. The main idea is that the vacuum expectation value of the singlet $\phi$ is not a constant but a function of temperature, where it becomes significant in the early universe and vanishes in the late universe. In contrast to the minimal axion assisted DW mechanism, the ensuing resonant oscillations enable an observable relic abundance associated with significantly reduced portal couplings $g_a$, evading the astrophysical constraints.
%%%%%%%%%%%%%%%%%%%%%%%%%%%%%%%%%%%%%%%%
\section{Production of sterile neutrino}
%%%%%%%%%%%%%%%%%%%%%%%%%%%%%%%%%%%%%%%%

Sterile neutrino is among the simplest solutions which generate the nonzero neutrino masses confirmed by oscillation experiments. The mixing between it and SM neutrino will then naturally give birth to the initial abundance of sterile neutrino. An accurate calculation of sterile neutrino production requires solving the evolution equation in the matrix density formalism, 
while in the collision-dominated period (which it's indeed the case of the production peak of sterile neutrino, i.e. $T\sim 0.1~\GeV$), the quasi-classical Boltzmann equation is a good approximation~\cite{Abazajian:2001nj}, 
\begin{eqnarray}
\left(\frac	{\partial}{\partial t} - Hp \frac	{\partial}{\partial p}\right)f_{\nu_s}(p,t) &\simeq& \frac{\Gamma_\alpha(p)}{2} \langle P_m(\nu_\alpha \leftrightarrow \nu_s;p,t) \rangle f_{\nu_\alpha}(p,t), \nonumber\\
\label{eq:BE}
\end{eqnarray}
here $f_{\nu_s}$ and $f_{\nu_\alpha}$ stand for the distribution functions of sterile neutrino and SM neutrino, $\Gamma_\alpha$ is the reaction rate where SM neutrino takes part in. $\langle P_m \rangle$ is the averaged converting probability between sterile neutrino and SM neutrino, expressed as~\cite{Volkas:2000ei,Lee:2000ej}
\begin{equation}
	\langle P_m(\nu_\alpha \leftrightarrow \nu_s;p,t) \rangle \simeq \frac{1}{2} \frac{\Delta^2 \sin^2 2\theta}{\Delta^2 \sin^2 2\theta + \Gamma_\alpha^2/4 + (\Delta \cos 2\theta - V)^2},
	\label{eq:cvtp}
\end{equation}
with $\Delta = m_s^2/2p$, the effective potential is shown as $V$, which characterizes the effects of SM neutrinos passing through the thermal background. The effective potential consists of contributions from SM electroweak interactions and any other new physics. For DW mechanism, there is only the SM electroweak contribution, which is approximated as $V_{\rm{SM}} \simeq - 14\pi/(45\alpha) (3-s_w^2)s_w^2 G_F^2 p T^4 \equiv -C_W^2 p T^4$~\cite{Notzold:1987ik,Alonso-Alvarez:2021pgy} for temperature between the mass of corresponding charged lepton and electroweak phase transition. The $\alpha$ stands for the fine structure constant, $s_w$ is the sine of Weinberg angle, and $G_F$ is the Fermi constant.

To obtain the relic abundance of sterile neutrinos, one can derive the equation for $r\equiv n_s/n_\alpha$, with $n_i$ denoting the number density of sterile (SM) neutrino when $i=s$ $(\alpha)$. The resulting equation reads
\begin{equation}
	\frac{d r}{d \ln a} = \frac{\gamma}{H} + r \frac{d \ln g_s^\star}{d \ln a}.
\end{equation}
Here entropy density conservation has been considered, and $g_s^\ast$ is the degrees of freedom with respect to entropy density. $\gamma$ on the right hand side is defined as
\begin{equation}
	\gamma \equiv \frac{1}{n_\alpha} \int \frac{d^3p}{(2\pi)^3}  \Gamma \langle P_m \rangle  f_{\nu_\alpha}.
\end{equation}
For constant $g_s^\ast$, $\gamma/H$ stands for the relative number of sterile neutrino, compared to SM neutrino, in each log-interval of $T$. Thus in the simple DW mechanism there are two free parameters, i.e., the mixing angle $\theta$ and the sterile neutrino mass $m_s$. The relic abundance of sterile neutrino will be determined once these two parameters are fixed. In Figure \ref{fig:dwadw}, we show the variation of $\gamma/H$ with temperature during a period of $[10^{-3}, 10]~\GeV$ in the DW mechanism (the red line). We choose the mixing angle as $\sin^22\theta=10^{-12}$, and the sterile neutrino mass is selected as $m_s = 10~\keV$. The production peak reaches $\gamma/H\sim 10^{-6}$ at temperature $\sim 0.2~\GeV$.

Such a framework is simple enough to produce sterile neutrino, just from SM neutrino oscillation.
However this mechanism is facing crises from various experiments. 
As the sterile neutrino are much heavier than SM neutrino, decay process of $\nu_s \to \nu_\alpha \gamma$ is inevitable, the decay rate is~\cite{Pal:1981rm}
\begin{equation}
	\Gamma(\nu_s\to \nu_\alpha \gamma) \simeq \sin^2 \theta \frac{9 G_F^2 \alpha m_s^5}{2^{11}\pi^4}.
\end{equation}
Hence the observation of X-ray~\cite{Watson:2011dw,Horiuchi:2013noa,Perez:2016tcq,Dessert:2018qih,Ng:2019gch} can put strong bounds on the mixing angle. 
Moreover, for fermionic DM, there is a lower mass limit due to the requirement that the phase-space density of DM does not exceed that of the degenerate Fermi gas~\cite{Tremaine:1979we}.
A bound of $m_{s} > 2~\keV$ is set from the analysis of DM phase space distribution in dwarf galaxies~\cite{Boyarsky:2008ju,Merle:2015vzu,Abazajian:2017tcc}. 
As can be seen in Figure~\ref{fig:adwfinal}, the X-ray observation has excluded almost the whole parameter space in DW mechanism. 
Considering the constraints of dwarf galaxies, there is no room for the DW mechanism to survive.

%%%%%%%%%%%%%%%%%%%%%%%%%%%%%%%%%%%%%%%%%%%%%
\section{The axion assisted DW mechanism}
%%%%%%%%%%%%%%%%%%%%%%%%%%%%%%%%%%%%%%%%%%%%%

We consider a new interaction of SM neutrino in an effective operator form
\begin{equation}
	-\mathcal{L}_a = g_a \left( \frac{\phi}{\Lambda}\right) \partial _\mu a ~\overline{\nu_\alpha} \gamma^\mu \gamma_5 \nu_\alpha,
	\label{eq:axionint}
\end{equation}
including the axion field $a$ with mass $m_a$, and a singlet $\phi$. 
The axial-vector interaction form is fixed by the axion shift symmetry. The inclusion of $\phi$ is to avoid severe constraints of neutrino oscillation observations. 
Scattering between axion and SM neutrino will modify the vacuum dispersion relation of neutrino, hence generating a Mikheyev-Smirnov-Wolfenstein matter effect~\cite{Wolfenstein:1977ue,Mikheyev:1985zog,Mikheev:1986wj}. In later discussions we can see that, without the inclusion of $\phi$, the coupling strength $g_a$ needed to give the correct DM relic density is in strong contradiction with the oscillation observations. 
The coupling strength matrix $g_a$ has a negative mass dimension $[M]^{-1}$. $\Lambda$ is the ultra-violate cutoff scale. 
Due to the large occupation number nature, the axion fields can be approximated in a classical wave form of $a(t,x) = a_0 \cos (m_a t - \vec p \cdot \vec x)$, with $a_0 = \sqrt{2\rho_a} /m_a$ and $\rho_a$ being the energy density of axions. Equation~\ref{eq:axionint} indicates that axions could be generated during the early stages of the universe. They could be produced thermally and maintained equilibrium with the SM bath. However, the resulting density is calculated to be significantly lower than that of cold dark matter~\cite{Graf:2010tv}. Consequently, they cannot be considered the primary component of dark matter. In our consideration, we treat sterile neutrino as a dominant component of dark matter, while for the axion we set $\rho_a$ as $1\%$ of the dark matter energy density.

Containing the singlet $\phi$ in the effective interaction is an important ingredient in our setup.
$\phi$ is a morphon field ~\cite{Croon:2020ntf} which develops a dynamically varied vacuum expectation value by interacting with Ricci scalar $R$ at the early stage of the universe, 
\begin{equation}
    V(\phi)= \frac12 m_\phi^2 \phi^2 - \frac{1}{2} \xi R \phi^2 + \frac14 \lambda_\phi \phi^4. 
\end{equation}
The coupling constant $\xi$ quantifies the interaction strength between the morphon field and curvature scalar. 
One common choice is $\xi=1/6$, which comes from the requirement of conformal invariance~\cite{Sonego:1993fw,Dowker:1970vu}. A comprehensive analysis of the temperature dependence of Ricci-scalar is carried out in~\cite{Caldwell:2013mox} for the time range from inflationary epoch to the present day. In our calculation, we consider the evolution of Ricci scalar from $T\sim 1$ MeV to $T\sim 10$ GeV. 
Thus, at high temperatures the scalar field develops a nonzero vacuum expectation value, 
\begin{align}
    v_\phi (T) = 
        \left\{
        \begin{array}{cl}
        \sqrt{ \frac{1}{\lambda_\phi} \left( \xi R(T) - m_\phi^2 \right)} \,, \ \ \ 
         &    \xi  R(T)  > m_\phi^2 \\[11pt]
        0       & \text{otherwise}\,.
        \end{array}
        \right. 
    \label{eq:vphi}
\end{align}
The scalar vacuum expectation value decreases with $T$  until it vanishes at low temperatures.
This is the main feature that it helps generate resonant oscillations in the early universe and evade constraints in the late universe, such as the neutrino oscillation bounds.

Now the additional contribution to the effective potential of SM neutrino can be obtained from the new introduced effective interaction in Equation~\ref{eq:axionint}, in the first-order approximation as
\begin{equation}
	V_a = g_a  \frac{v_\phi(T)}{\Lambda} \left(- \partial_0 a + \vec \partial a \cdot \vec p_\nu /|\vec p_\nu|\right).
	\label{eq:va}
\end{equation}
The derivative on space part could be neglected as the velocity of axion in the Milky Way is of $\mathcal{O}(10^{-3})c$. Then the effective potential is further approximated as
\begin{equation}
	V_a \approx g_a  \frac{v_\phi(T)}{\Lambda} \sqrt{2\rho_a} \sin (m_a t),
	\label{eq:simva}
\end{equation}
where the plane-wave form of the axion field has been taken.

\begin{figure}[!htbp]
    \centering
    \includegraphics[width=0.45\textwidth]{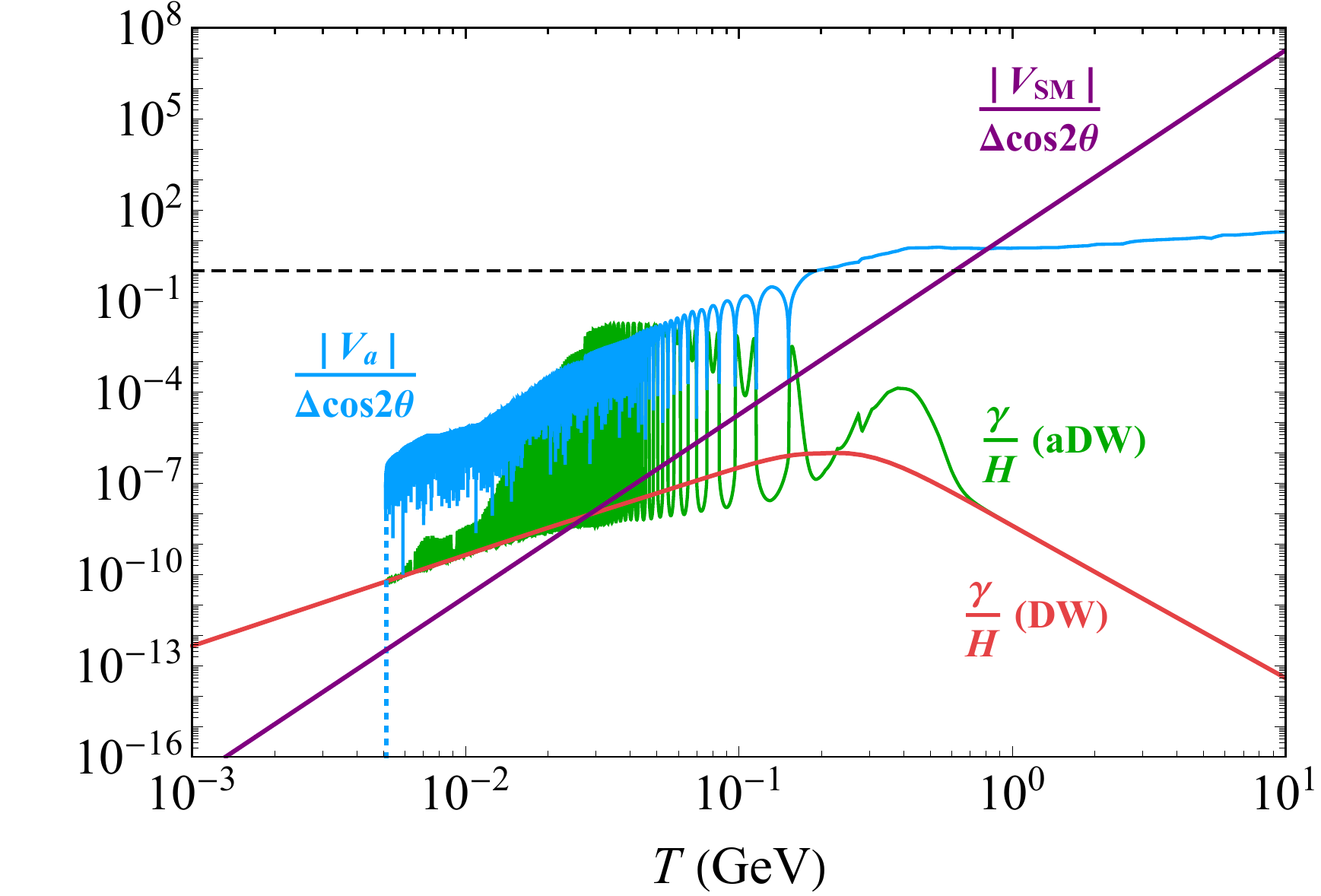}
    \caption{Production of sterile neutrino in the DW and aDW mechanisms. The red and green lines stand for the variation of $\gamma/H$ with temperature $T$ in DW and aDW mechanisms, respectively. The parameters are set as $\sin^2 2\theta = 10^{-12}$, $m_s = 10~\keV$, $m_\phi = 10^{-15}~\eV$, $\xi = 1/6$, $m_a = 10^{-10}~\eV$,
and a combination of $g_a/(\sqrt{\lambda_\phi} \Lambda)$ is set to $10^{14}~\eV^{-2}$. The blue and purple lines show relative size of the SM potential and the new potential to $\Delta \cos 2\theta$.}
    \label{fig:dwadw}
\end{figure}

We have mentioned above that the SM weak interactions contribute a negative thermal potential $V_{\rm{SM}} = - C_W^2 p T^4$. 
While the new contribution could be either positive or negative, depending on the axion mass and time (or temperature).
%depending on the sine function $\sin(m_a t)$. %
Therefore, in principle there exists cancellation between the contributions from SM and the new interaction. 
%in $\Delta \cos 2\theta - V_{\rm{SM}} - V_a$, 
It would substantially enlarge the averaged converting probability $\langle P_m \rangle$, and resonantly produce the sterile neutrino dark matter. As a result, the mixing angle could be much smaller than that in the original DW mechanism. 
This can help to release the strong tension against the X-ray observations. 
%which suffers severe constraints from X-ray observations, 
The cancellation condition is explicitly written as
\begin{equation}
    \Delta \cos 2\theta - V_{\rm{SM}} - V_a = 0.
    \label{eq:res}
\end{equation}
We firstly give a rough estimation to show the incapability of the minimal interaction. Actually the minimal interaction would arise when $\phi$ develops a vacuum expectation value, under the condition $\xi R(T) > m_\phi^2$. Specifically we label the minimal interaction as $g_a^\prime \partial_{\mu}a \overline{\nu_\alpha}\gamma^{\mu}\gamma_5 \nu_\alpha$, here $g_a^\prime$ relates to $g_a$ as $g_a^\prime =  g_a v_\phi(T)/\Lambda$. The generated potential is in a form of $g_a^\prime \sqrt{2\rho_a} \sin(m_a t)$.
The former two terms in Equation~\ref{eq:res} have different dependencies on temperature, i.e. $\Delta \cos 2\theta \propto T^{-1}$ and $V_{\rm{SM}} \propto T^5$ with setting $p=T$ for simplicity. Thus the lowest value of the sum is estimated as $\sim 4.1(m_s/10~\keV)^{5/3}~\eV$. 
We find that the coupling strength $g_a^\prime$ needed for the effective potential, to have a considerable cancellation with the former two terms in Equation~\ref{eq:res}, is much above the upper limits from the oscillation experiments. 
For instance, $g_a^\prime$ should be no less than $\sim 10^4~\eV^{-1}$ for the configuration we considered about dark matter components. However it's much larger than the upper bounds from oscillation experiments~\cite{Huang:2018cwo}. Taking DUNE as an example, the upper bounds of $g_a^\prime$ lie in $[3\times 10^{-12}, 3\times 10^{-9}]~\rm{eV}^{-1}$ for $m_a$ in $[10^{-22}, 10^{-9}]~\rm{eV}$~\cite{Huang:2018cwo}.  Thus if the minimal interaction were to persist until the present epoch, it would significantly impact neutrino oscillations, imposing restrictions both on the parameter $g_a^\prime$ and, consequently, on $g_a$. However, the oscillation limitations would not appear in our case, since the introduced $\phi$ has a temperature-dependent vacuum expectation value(see Equation \ref{eq:vphi}). The selected $m_\phi$, i.e. $m_\phi=10^{-15}~\eV$, would make the vacuum expectation value of $\phi$ to be vanished at around $5~\MeV$. Thus, at the present time, both the minimal interaction and the effective potential in \ref{eq:va}, \ref{eq:simva} are absent, as $v_\phi$ is vanished. The impact on oscillations is also eliminated. As a result, we can get rid of the oscillation constraints.
Another simple interaction form between SM neutrino and scalar is $\phi \bar{\nu} \nu$. From Ref.~\cite{Ge:2018uhz} we know it only results in a neutrino mass correction, and is irrelevant to the resonance oscillation. Neutrino coalescence into axion and $\phi$ could take place from the effective interaction \ref{eq:axionint}, it may affect the cooling process in supernovae. In order to avoid any discrepancies with the observation of SN1987a, it is necessary for the coupling strength to guarantee that the amount of energy carried away by the new particles does not exceed the value permitted by the neutrino-only scenario. This condition is referred to as the ``Raffelt criterion"~\cite{Raffelt:1996wa}.  In essence, this requirement could be met if $g_a/\Lambda$ is much less than the strength of weak interaction, i.e. $G_F$. 
The resonant production of sterile neutrino and constraints from supernovae cooling necessitate an exceedingly tiny value for $\lambda_\phi$. For instance, with typical parameters as shown in Figure \ref{fig:dwadw}, the ratio $g_a/(\sqrt{\lambda_\phi} \Lambda)$ is set to $10^{-14}~\rm{eV}^{-2}$. To satisfy the condition that $g_a/\Lambda$ is much smaller than $G_F$, it implies that $\lambda_\phi$ must be much less than $10^{-74}$.
Such a tiny $\lambda_\phi$ might seem unnatural, but one potential solution is the Clockwork mechanism~\cite{Park:2018kst}. By introducing a series of scalar gear fields $\phi_i(i=1...N)$, a scalar potential can be constructed as~\cite{Park:2018kst}
\begin{equation}
    V_{\rm{gear}} \supset \frac{m^2}{2} \sum_{i=1}^N (\phi_{i+1} - q \phi_i)^2 + \frac{\lambda}{4} \phi_1^4.
\end{equation}
The iterative relationships between these gear fields, determined by the Euler-Lagrange equation of motion, result in $\phi_1$ being exponentially suppressed as $\phi_1 \sim q^{-N} \varphi_{(0)}$, where $\varphi_{0}$ is defined as $|\partial_\mu \varphi_{0}|^2 \equiv \sum_i^{N+1} |\partial_\mu \phi_i|^2$. This allows us to express the quartic interaction in terms of an effective coupling $\lambda_\phi$, which can be naturally generated from an $\mathcal{O}(1)$ coupling $\lambda$ via $\lambda_{\phi} \sim q^{-4N}\lambda$, for $q>1$. For instance, choose $q=5$, $N=28$ and $\lambda$ at $\mathcal{O}(1)$, the resultant $\lambda_\phi \sim 10^{-80}$.

%It is easy to fulfill this condition when taking into account the resonant production of sterile neutrinos, since the selection of $\lambda_\phi$ offers some degrees of flexibility.

To reveal the resonance effects, we depict the variation of the two potentials, as functions of temperature in Figure~\ref{fig:dwadw}. In practice, we normalize these two terms by $\Delta \cos 2\theta$, and momentum is set to $p=T$. 
The purple line indicates the relative {magnitude of} SM potential, 
and the blue line stands for the relative {magnitude of} the new potential $V_a$. 
The black dashed line represents the normalization factor $\Delta \cos 2\theta$.

Without considering large deviations between the sterile neutrino momentum and temperature, 
it can be seen from Figure~\ref{fig:dwadw} that $\Delta \cos 2\theta$ is dominant in the region of $T\lesssim 0.2~\GeV$, 
while the SM potential $|V_{\rm{SM}}|$ is dominant in the region of $T\gtrsim 0.8~\GeV$. 
The new potential $V_a$ oscillates with temperature and terminates at $T \sim 5~\MeV$ because the vacuum expectation value  vanishes at lower temperatures. 
As the temperature increases, the oscillation frequency becomes smaller since the relation between time and temperature is $t\propto T^{-2}$. 
The maximum value that can be reached by the oscillated $V_a$ gradually approaches and exceeds $\Delta \cos 2\theta$. 
Therefore, as the temperature increases, the cancellation between the three terms is expected to be more pronounced, until $|V_{\rm{SM}}|$ increases so that SM contribution becomes dominant.

In Figure~\ref{fig:dwadw}, we also show the dependence of $\gamma/H$ on the temperature for both DW (red line) and aDW (green line) mechanisms. The aDW scenario will go back to DW for the high and low temperatures. 
For the intermediate temperature, the cancellation leads to an enhancement on $\gamma/H$ in the aDW mechanism, whose magnitude is up to $\sim 6$ orders higher in the comparison of the two scenarios. 
Thus the production of sterile neutrino dark matter is expected to be greatly enhanced for aDW.

\begin{figure}
    \centering
\includegraphics[width=0.45\textwidth]{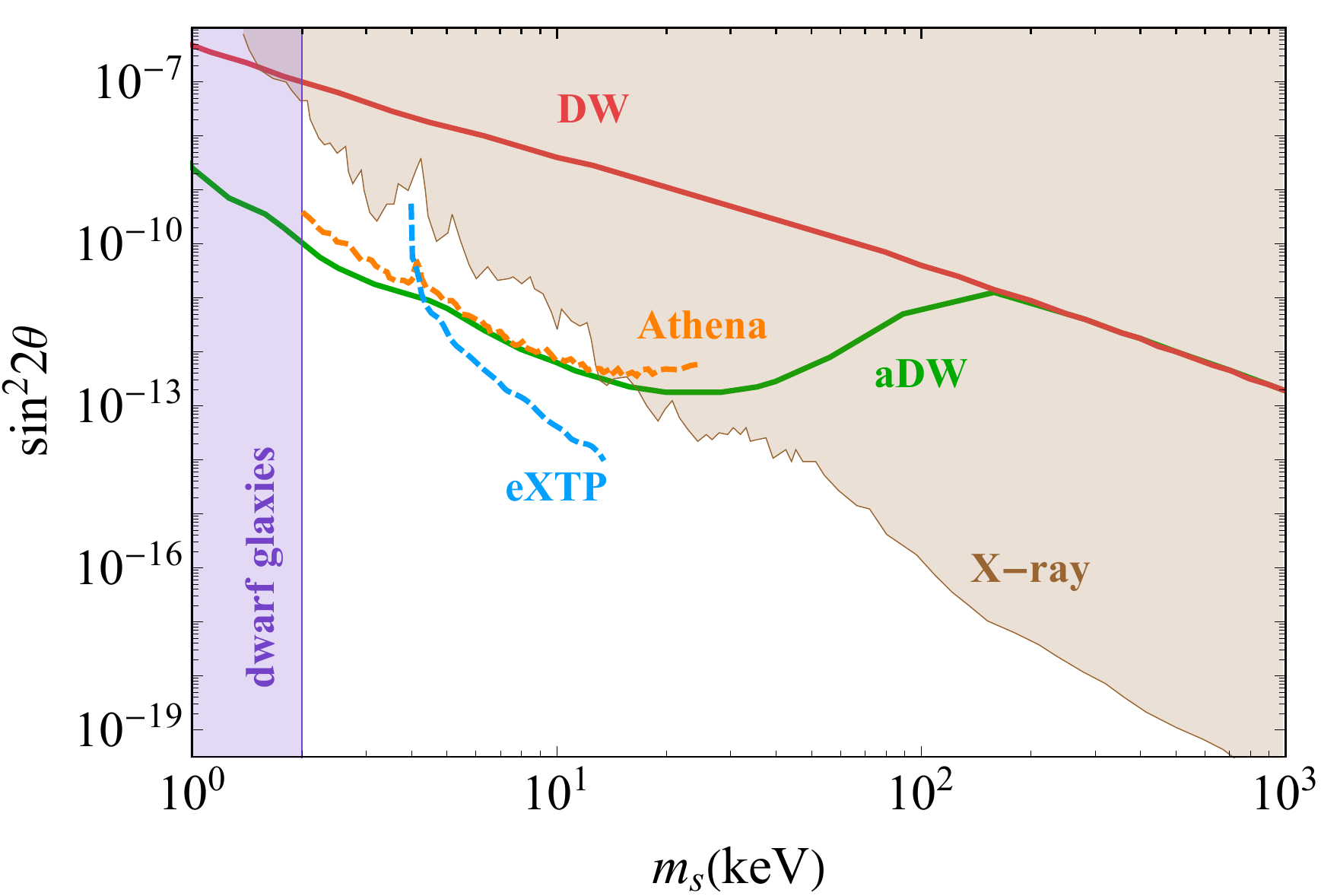}
    \caption{Correct relic density of sterile neutrino DM as functions of mass and mixing angle. The red line represents the proper parameter space in the  traditional Dodelson-Widrow mechanism (DW), while the green line stands for that in the aDW mechanism (aDW). The choices of model parameters are $m_\phi = 10^{-15}~\eV$, $\xi = 1/6$, $m_a = 10^{-10}~\eV$, and $g_a/(\sqrt{\lambda_\phi} \Lambda)=10^{14}~\eV^{-2}$. Constraints from X-ray observation are depicted in brown region, and lower bound on mass of sterile neutrino from dwarf galaxies is shown by purple region. The detecting capabilities of two forthcoming experiments, i.e. Athena and eXTP, are showed in orange and blue dashed lines respectively.}
    \label{fig:adwfinal}
\end{figure}

{Base on the resonance effects, the sterile neutrino could be produced with a much smaller mixing angle.}
In Figure~\ref{fig:adwfinal}, we show the correct relic density in ($m_s$, $\sin^2 2\theta$) plane. 
The green line stands for the available parameter space in aDW scenario, and the red line represents the traditional DW mechanism. The most stringent constraints come from the X-ray observation and phase space distribution of dwarf galaxies, which are shown by the brown and purple excluded regions. The most parameter space that predicted by DW mechanism have been excluded. 
In our scenario containing the new interactions, the values of the mixing angle can be much lower comparatively, meanwhile providing the correct amount of relic abundance. 
For the parameters we have selected, sterile neutrinos with mass in $[2, 12]~\keV$ can satisfy all the astrophysical constraints. 
In Figure~\ref{fig:adwfinal}, we also show the detecting capabilities of two forthcoming experiments, i.e. Athena~\cite{Neronov:2015kca} and  eXTP~\cite{Zhong:2020wre,Malyshev:2020hcc}. 
We see that they could detect some parameter region that is more stringent than the X-ray constraints, and can be used to test the feasibility of our scenario.

%%%%%%%%%%%%%%%%%%%%%%%%%%%%%%%%%%
\section{Conclusion}
%%%%%%%%%%%%%%%%%%%%%%%%%%%%%%%%%%

Sterile neutrino dark matter, generated via the mixing with SM neutrino in DW mechanism, 
suffers severe conflict between its relic density and the astrophysical observations. 
X-ray constraints excluded the most parameter space of this scenario. 
To release this tension, we have introduced an effective axial-vector interaction between axion, SM neutrino, and a singlet in this work.
The new interactions bring a new contribution to SM neutrino effective potential, 
then the cancellation occurs between this new contribution and SM one.
The sterile neutrino thus can be resonantly produced with a significant enhancement factor. 
Around the typical DW-peak temperature, the resonance enhancement could produce sterile neutrino with an amount of several orders higher than DW mechanism. 
Such that the much smaller mixing angles can satisfy the stringent astrophysical constraints and provide the right abundance of the relic DM at the same time. 
An inevitable consequence of new interaction with neutrino is the modification of the neutrino oscillation. Since the vacuum expectation value of the new scalar dynamically varies in temperature, the new effective potential would vanish at the current temperature. So it's safe to ignore it in our case.

%%%%%%%%%%%%%%%%%%%%%%%%%%%%%%%%%%
\section*{Acknowledgements}
%%%%%%%%%%%%%%%%%%%%%%%%%%%%%%%%%%

We thank Guo-yuan Huang for helpful discussions. 
This work was supported by the National Natural Science Foundation of China under Grants Nos. 12005180, 12275232, by the Natural Science Foundation of Shandong Province under Grants No. ZR2020QA083, and by the Project of Shandong Province Higher Educational Science and Technology Program under Grants No. 2019KJJ007.

%\appendix

\bibliography{refs.bib}
\bibliographystyle{apsrev4-1}
\newpage
\end{document}